\documentclass[epj]{svjour}
\usepackage{latexsym}
\usepackage{graphics}
\begin{document}
\title{Coupled Aggregation and Sedimentation Processes: \\Three
Dimensional Off-Lattice Simulations}
\author{R. Leone\inst{1} \and G. Odriozola\inst{1} \and L. Mussio\inst{1}
\and A. Schmitt\inst{2} \and R. Hidalgo-\'{A}lvarez\inst{2}
\thanks{\emph{e-mail:} rhidalgo@ugr.es}%
}                     
%
%
\institute{Departamento de Qu\'{\i}mica F\'{\i}sica y
Matem\'{a}tica, Facultad de Qu\'{\i}mica\\ Universidad de la
Rep\'{u}blica, 11800 Montevideo, Uruguay \and Departamento de F\'{\i}sica Aplicada;
Facultad de Ciencias; Universidad de Granada;\\
Campus de Fuentenueva; E-18071 Granada; Spain}
\date{Received: date / Revised version: date}
%
\abstract{Coupled aggregation and sedimentation processes were
studied by means of three dimensional computer simulations. For
this purpose, a large prism with no periodic boundary conditions
for the sedimentation direction was considered. Furthermore, three
equally sized and mutually excluded regions were defined inside
the prism, a top, a middle and a bottom region. This allows to
study the time evolution of the cluster size distribution and the
cluster structure separately for each region. The mass
distribution profile and the center of mass position were also
accessed as a function of time. For the bottom region, the effects
of the sediment formation on the kinetics of growth and on the
cluster structure were clearly observed. The obtained results not
only agree with the experimental data obtained by Allain
et.~al.~\cite{Allain95} and with the simulations made by Gonzalez
\cite{gonzalez01} but also allow to gain further insight in the
details.}

%
\maketitle
\section{Introduction}
\label{intro} To a larger or smaller extent, all real aggregation
processes are influenced by the presence of an external
gravitational field. Hence, the aggregation phenomena are usually
followed by sedimentation or, in many cases, both are present
simultaneously. Some natural examples are the delta formation at a
river estuary and the settling of bacteria clusters in quiet water
\cite{lyklema91}. As technological examples one may cite water
treatment for human consumption (clarifying), effluents treatment,
and a large number of precipitation techniques employed by the
chemical industry \cite{hunter87}.

Pure irreversible aggregation processes have been well studied and
are described in the literature. The first equation for describing
the aggregation kinetics of diluted systems was given by
Smoluchowski in the early 1900s \cite{smol16,smol17}. This
equation defines an infinite two-dimensional matrix of kinetic
rate constants, known as kernel, which accounts for the physical
characteristics of the system. The time evolution of the cluster
size distribution is obtained by solving the Smoluchowski's rate
equation for a given kernel. On the other hand, much information
about the aggregation processes is also contained in the clusters
structure. Here, Smoluchowski's rate equation does not provide
useful information and hence, other techniques such as light
scattering experiments and simulations are required for its study.
Significant contributions in these fields are provided by Lin,
Weitz and coworkers (light scattering experiments)
\cite{lin89,lin90pcm,weitz84a} and Meakin, Family and Gonzalez
(simulations)
\cite{meakin83,meakin87,family85,gonzalez93,gonzalez96}.

Pure sedimentation phenomena have also been well studied
\cite{russel89}. Here, an important issue is the variation of the
settling velocity with the volume fraction
\cite{Batchelor82a,Batchelor82b}. When the concentrations become
large enough, the reverse flow of fluid necessary to compensate
the volumetric flow of particles plus the associated fluid
contribute to decrease the Stokes velocity (this is known as
backflow effect). Furthermore, since the volume fraction becomes
larger when moving to the flask bottom, an osmotic pressure
appears opposed to the gravitational field. Finally, for larger
volume fractions the inter-particle distances shorten and hence,
the role of hydrodynamic forces and inter-particle interactions
become important.

Although aggregation and sedimentation phenomena are closely
connected, there is not an extended literature dealing with them
simultaneously. This is, at least partially, due to the
mathematical difficulties, which appear when following a formal
analytical treatment \cite{Stanley01}. Consequently, computer
simulations become a useful alternative tool for studying and
predicting the behavior of real aggregation-sedimentation systems.
Recently, Gonzalez and Leyvraz worked on computer simulations for
elucidating the experimental results found by Allain et. al.
\cite{Allain95,Allain96}. They found that increasing the intensity
of the external field leads to an increase in the cluster fractal
dimension. Their model, however, did not consider cluster
deposition (only a sedimentation velocity was added to the
Brownian motion) \cite{gonzalez96i}. Later, the simulations were
improved by considering a cubic lattice of size L, with periodic
boundary conditions in the three spatial directions (as in the
previous model), but now taking the clusters out of the cubic box
with a probability related to the total distance that they had
moved downwards (see ref. \cite{gonzalez01} for details). Although
this model does not consider the mass distribution dependence on
height, it was useful for explaining an increase of the initial
particle concentration required to have gelation when
sedimentation is present.

In this paper, we further study the coupled aggregation and
sedimentation phenomena by simulations, but now considering a
large prism with no periodical boundary conditions for the
sedimentation direction (this requires a macroscopic prism
height). For the other two horizontal directions, we still impose
periodical boundary conditions in order to represent, in an
average way, a portion of the whole system. This allows us to
study not only average quantities but also their change as a
function of height. Furthermore, it was possible to observe and
study the sediment formation.

\section{Simulation method}
\label{simul} The simulation processes were carried out off
lattice on a square section prism of side $L$ and height $H$.
Inside, $N_0$ identical hard spheres of radius $a$ were randomly
placed avoiding overlapping among them. Since aggregation and
sedimentation processes are simulated, there are two contributions
to the movement. One corresponds to the Brownian motion and the
other to the Stokes sedimentation velocity. As the simulations
were done for very dilute systems, backflow and hydrodynamic
forces were not taken into account. This assumption will have to
be checked since local high concentrations may appear somewhere in
the prism. The time step was fixed by the relationship
\begin{equation}
t_0=l_B^2/(6D_1)
\end{equation}
where $D_1$ $=$ $k_BT/(6\pi\eta a)$ is the monomer diffusion
coefficient, $k_BT$ is the thermal energy, $\eta$ is the solvent
viscosity and $l_B$ is the Brownian length step. Hence, monomers
are always moved $l_B$ in a random direction plus the Stokes
contribution
\begin{equation}
l_S=v_St_0=l_B^2Pe/6a
\end{equation}
where $v_S$ $=$ $2(\rho-\rho_0)ga^2/(9\eta)$ is the monomeric
Stokes velocity, $\rho$ is the particle density, $\rho_0$ is the
fluid density, $g$ is the earth gravitational constant and $Pe$
$=$ $4\pi a^4(\rho-\rho_0)g/k_BT$ is the Peclet number. For
$i$-sized aggregates, we assume the relationship $r_g$ $=$
$ai^{1/df}$ for the radius of gyration and therefore the diffusion
coefficient $D_i$ $=$ $k_BTi^{-1/df}/(6\pi\eta a)$ and the Stokes
velocity $v_S$ $=$ $2(\rho-\rho_0)ga^2i^{1-1/d_f}/9\eta$. Here, it
was implicitly assumed that the hydrodynamic fractal dimension and
the cluster fractal dimension are equal. Now, in order to account
for the Brownian motion, we move the aggregates $l_B$ in a random
direction only when a random number, $\xi$, uniformly distributed
in [0,1], is less than the ratio between the aggregate diffusion
coefficient and the monomer diffusion coefficient, i.~e.\ when
$\xi$ $<$ $i^{-1/df}$. In case that the Brownian movement is
refused, the corresponding Stokes contribution
\begin{equation}
l_S=v_St_0=l_B^2Pei^{1-1/d_f}/6a
\end{equation}
is accumulated in a memory place, associated with the particular
cluster, for being considered in the following time intervals.
Once a Brownian contribution is accepted, the accumulated Stokes
contribution is added to the cluster motion and the corresponding
accumulated memory is reset to zero. Furthermore, if a given
accumulated Stokes contribution exceeds a Brownian step, $l_B$,
then the aggregate is moved downwards and the accumulated memory
is also reset to zero, thus avoiding a large total step length.
The algorithm minimizes the times that clusters are moved, which
is the most time consuming contribution to the execution time.
After any movement, the regional configuration is checked for
overlaps. In case that any overlap is found, it is corrected by
placing the recently moved cluster in touch with the other to form
a new aggregate, i.~e.~diffusion limited cluster aggregation
(DLCA) conditions are imposed.

Periodic boundary conditions were established for the two non
sedimentation directions, $x$ and $y$. Hence, the system may be
understood as a small portion of a macroscopic one. However, for
the sedimentation direction, $z$ (we define the prism bottom as
$z$ $=$ $0$ and the prism top as $z$ $=$ $1$), no periodic
boundary condition was imposed in order to naturally obtain a
change in properties with the prism height and a cluster deposit
at the prism bottom. This forces us to use a high prism since the
height of the macroscopic system coincides with its vertical
length. This could be done while keeping a reasonable total number
of particles, i.~e.~working with diluted systems. As the system
symmetry was broken for the $z$ direction, we look for this effect
on the cluster populations and on the cluster structures.

For this purpose, the system volume was arbitrarily divided in
three equally sized and mutually excluded regions: a top, a middle
and a bottom region. Hence, we may define the weight average
cluster size for each region as
\begin{equation}
n_w|_h = \frac{\sum_i i^2n_i|_h}{\sum_i i n_i|_h}
\end{equation}
where the symbol $|_h$ $=$ $top$, $middle$ or $bottom$ refers to
the corresponding region and $n_i$ is the number of $i$-sized
clusters. Furthermore, the vertical position of the system center
of mass is also defined as
\begin{equation}
z_{cm} = \frac{\sum_l i_l z^{cm}_l}{\sum_l i_l}
\end{equation}
where $i_l$ is the size of cluster $l$ and $z^{cm}_l$ is the
vertical position of the center of mass of cluster $l$.

\begin{figure}
\resizebox{0.47\textwidth}{!}{\includegraphics{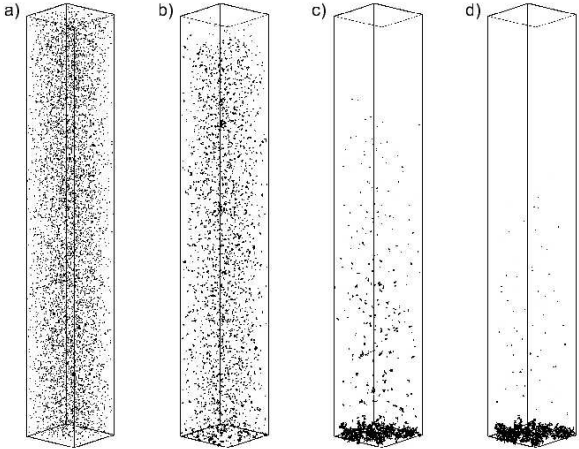}}
\caption{\label{fig1} Three dimensional representation of a
simulated system settling under $Pe$ $=$ $0.1$. The prism
dimensions are $H$ $=$ $1500a$ and $L$ $=$ $250a$ ($H$ $=$ $1500a$
was established instead of $H$ $=$ $10000a$ for an easier
representation). The images were captured at times 0.1 s a), 112 s
b), 630 s c) and 1120 s d). }
\end{figure}

The cluster structure is studied by calculating the following
radii of gyration every time a new cluster is formed
\begin{equation}\begin{array}{l@{\;=\;}l}
 r_{gx}& \sqrt{\frac{1}{n}\sum_i(\vec{r_i}.\vec{r_i}-x_i^2)} \\
 r_{gy}& \sqrt{\frac{1}{n}\sum_i(\vec{r_i}.\vec{r_i}-y_i^2)} \\
 r_{gz}& \sqrt{\frac{1}{n}\sum_i(\vec{r_i}.\vec{r_i}-z_i^2)} \\
 r_{g}& \sqrt{\frac{1}{n}\sum_i(\vec{r_i}.\vec{r_i})} \\
\end{array}
\end{equation}
where $\vec{r}_i$ is the distance between the particle $i$ and the
cluster center of mass and $x_i$, $y_i$ and $z_i$ are its
components. Those radii of gyration allow us to evaluate the
average ratios $\langle r_{gx}\rangle / \langle r_{gz}\rangle |_h$
and $\langle r_{gy}\rangle / \langle r_{gz}\rangle |_h$ and to
obtain three cluster fractal dimensions from $r_g|_h$ $=$
$ai^{1/d_f|_h}$ (considering only clusters containing more than 15
particles). The former quantities account for the average shape of
the clusters, i.~e.\ $\langle r_{gx}\rangle / \langle r_{gz}
\rangle$ $>$ $1$ indicates that the clusters tend to be elongated
in the $z$ direction and, on the contrary, $\langle r_{gx}\rangle
/ \langle r_{gz} \rangle$ $<$ $1$ points to wider structures.
$\langle r_{gx}\rangle / \langle r_{gz} \rangle$ and $\langle
r_{gy}\rangle / \langle r_{gz} \rangle$ should be similar but not
equal due to statistical fluctuations and so, their difference
represents an estimation of their uncertainty.

It should be noted that the cluster motion depends on the fractal
dimension, which is not known a priori. This forces us to estimate
$d_f$ and to iterate the simulations whenever the introduced $d_f$
differs from the obtained $d_f$ in more than $0.05$.

The parameters employed for the simulations are the following: a
monomer radius $a$ $=$ $315$ $nm$, a Brownian step length $l_B$
$=$ $a/2$, a prism height $H$ $=$ $10000a$, a section side $L$ $=$
$250a$ and a particle volume fraction of $\phi$ $=$
$6.7\times10^{-5}$ ($N_0$ $=$ $10000$). The particles were
considered to be dispersed in water at $20^0$ C. Different runs
were performed for different Peclet numbers. In real experiments,
a large change of the Peclet number may be achieved by changing
the strength of the external field by centrifugation. Another
possibility for realizing small changes of the Peclet number
consists in using particles with different densities.

\section{Results}
\label{results}
\subsection{Overview}

Figure \ref{fig1} was constructed to give an overview of the
coupled aggregation and sedimentation processes. It shows a three
dimensional representation of a system settling under $Pe$ $=$
$0.1$. The images were captured from simulations at times $0.1$ s
a), $112$ s b), $630$ s c) and $1120$ s d). In order to easily
represent the images, here we set $H$ $=$ $1500a$ instead of $H$
$=$ $10000a$, which is the value employed for the study. As can be
seen for the beginning of the process, figure \ref{fig1} a) shows
a uniformly distributed and monodisperse system. After a few
seconds, figure \ref{fig1} b), the system aggregates forming a
wide cluster size distribution. As imposed, the larger clusters
move faster downwards and hence, they are mostly seen close to the
prism bottom. As the times goes on, those clusters start forming
the sediment while the dispersion clarifies by loosing mass. The
first settling aggregates that arrive to the prism bottom continue
their Brownian motion although they are not allowed to move
further downwards. Hence, their movement is almost restricted to
two dimensions, since their weight makes difficult their upward
motion. Consequently, they collide among each other forming the
sediment. In the following time intervals, the system looks like
the one shown in figure \ref{fig1} c). Here, it is observed that
the mass concentration depends on the distance from the bottom of
the prism. At the top, the system is almost clear while the
dispersion looks more concentrated at the bottom. Furthermore, a
large sediment is also observed. It gains mass from every closed
enough settling cluster. On the other hand, the number of clusters
remaining in the dispersion strongly decreases and the aggregation
rate of the dispersion slows down. Consequently, aggregation
becomes highly improbable for those small remaining clusters and a
long time is needed for their deposition. This situation is
clearly seen in figure \ref{fig1} d).

It should be pointed out that this first overview is in good
agreement with the experimental observations made by Allain
et.~al.~for calcium carbonate colloidal suspensions
\cite{Allain95}. Their observations were performed experimentally
in a 800 mm high cell. Furthermore, they emphasize that the
different phases of settling cannot be identified separately for
smaller cells. This is not surprising since their experiments were
characterized by a Peclet number close to $10^{-5}$.

\subsection{Further inside in the details}

\begin{figure}
\resizebox{0.47\textwidth}{!}{\includegraphics{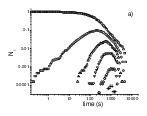}}
\resizebox{0.47\textwidth}{!}{\includegraphics{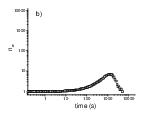}}
\caption{\label{fig2} a) Time evolution of the normalized
cluster-size distribution for the upper portion of the system, $h$
$=$ $top$, and $Pe$ $=$ $0.1$. The points are grouped in
logarithmically spaced intervals (({\tiny $\Box$}) monomers,
($\circ$) 2 and 3-mers, ({\tiny $\bigtriangleup$}) 4 to 8-mers,
({\tiny $\bigtriangledown$}) 9 to 18-mers, ($\diamond$) 19 to
38-mers and ({\tiny $+$}) 39 to 88-mers. b) The corresponding
weight average cluster size, $n_w$.}
\end{figure}

\begin{figure}
\resizebox{0.47\textwidth}{!}{\includegraphics{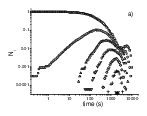}}
\resizebox{0.47\textwidth}{!}{\includegraphics{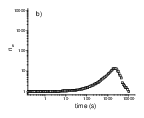}}
\caption{\label{fig3}a) Time evolution of the normalized
cluster-size distribution for the middle portion of the system,
$h$ $=$ $middle$ and $Pe$ $=$ $0.1$. The points are grouped in
logarithmically spaced intervals (({\tiny $\Box$}) monomers,
($\circ$) 2 and 3-mers, ({\tiny $\bigtriangleup$}) 4 to 8-mers,
({\tiny $\bigtriangledown$}) 9 to 18-mers, ($\diamond$) 19 to
38-mers, ({\tiny $+$}) 39 to 88-mers and ({\tiny $\times$}) 89 to
200-mers). b) The corresponding weight average cluster size,
$n_w$.}
\end{figure}

\begin{figure}
\resizebox{0.47\textwidth}{!}{\includegraphics{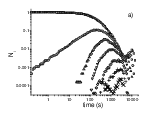}}
\resizebox{0.47\textwidth}{!}{\includegraphics{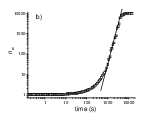}}
\caption{\label{fig4} a) Time evolution of the normalized
cluster-size distribution for the bottom portion of the system,
$h$ $=$ $bottom$ and $Pe$ $=$ $0.1$. The points are grouped in
logarithmically spaced intervals (({\tiny $\Box$}) monomers,
($\circ$) 2 and 3-mers, ({\tiny $\bigtriangleup$}) 4 to 8-mers,
({\tiny $\bigtriangledown$}) 9 to 18-mers, ($\diamond$) 19 to
38-mers, ({\tiny $+$}) 39 to 88-mers and ({\tiny $\times$}) 89 to
200-mers). b) The corresponding weight average cluster size,
$n_w$.}
\end{figure}

Figure \ref{fig2} a) shows the time evolution of the cluster-size
distribution for the upper portion of the system, $h$ $=$ $top$,
aggregating under $Pe$ $=$ $0.1$. As expected, the first stage of
the aggregation process evolves in time similarly to the DLCA
regime, i.~e. the sedimentation effects are not yet important.
Once the oligomer concentration becomes higher, the subsystem
starts to loose mass due to sedimentation and so, the formation of
larger clusters is biased. Furthermore, the characteristic bell
shape evolutions of oligomers narrow since they disappear not only
by reaction but also by leaving the subsystem. When comparing the
obtained time evolution of the monomer concentration with the one
corresponding to the DLCA regime, differences are also observed
for the later stages. For the pure DLCA regime, the monomer
concentration decays faster than in the case where sedimentation
effects are considered. The reason for this is that a larger
cluster concentration remains in the subsystem and hence, a higher
aggregation velocity is achieved. Moreover, since for DLCA
processes the monomeric aggregation rate constants are largest,
their curve crosses the curves for the larger clusters leading to
a smaller monomer concentration \cite{PRE00}. This is not observed
when sedimentation is present.

The corresponding weight average cluster size, $n_w$, for $h$ $=$
$top$ and $Pe$ $=$ $0.1$ is shown in figure \ref{fig2} b). Again,
at the beginning of the processes $n_w$ evolves in time similarly
to the DLCA regime. As time goes on, two effects tends to change
its evolution. On the one hand, the different sedimentation
velocities of clusters leads to an increase in their collision
frequency and consequently, the average $n_w$ tends to increase
its rate of change. On the other hand, the larger aggregates exit
the subsystem more often than the smaller ones tending to decrease
the average $n_w$. For a given time, the latter effect prevails
over the former and so, the average cluster size peaks. For larger
times, this situation makes the average $n_w$ to monotonically
decrease until monomers are the only species in the subsystem.

The time evolution of the cluster size distribution and the
corresponding average, $n_w$, for the second region, $h$ $=$
$middle$, are shown in figure \ref{fig3}. Although the evolutions
are similar than those shown for the upper region of the prism,
some differences are found. Since this subsystem also gain mass
from the upper region while loosing it towards the bottom, it is
capable to develop larger structures and to hold a large number of
clusters. This is clearly seen in figure \ref{fig3} a) where a
large number of $39$ to $88$-mers is shown and even a few $88$ to
$200$-mers appear. Furthermore, the weight average cluster size
peaks for larger times and reaches a higher value (compare figure
\ref{fig3} b) with  \ref{fig2} b)). Finally, it is observed that
the monomer population becomes almost constant at approximately
$t$ $=$ $4000$ $s$, which is the same time at which the upper
subsystem has almost lost the remaining monomers. This fact is
explained by means of the monomeric transfer from the upper region
towards the middle subsystem.

\begin{figure}
\resizebox{0.47\textwidth}{!}{\includegraphics{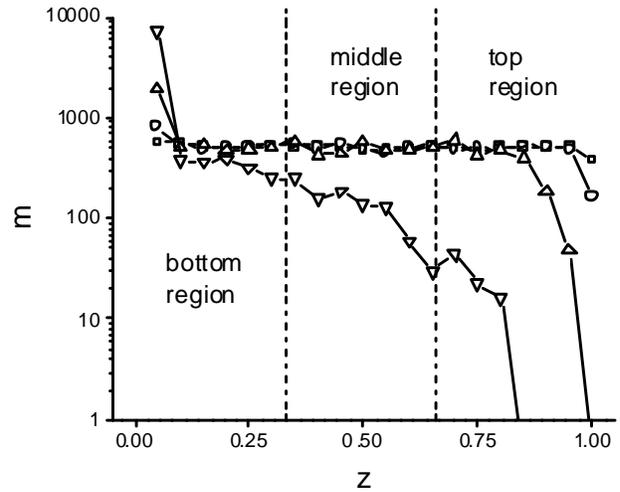}}
\caption{\label{fig5} Mass distribution profile, $m(z)$, along the
sedimentation direction, $z$, represented in log-normal axes for
different times. The data were obtained for $Pe$ $=$ $0.1$. The
symbols {\tiny $\Box$}, $\circ$, {\tiny $\bigtriangleup$} and
{\tiny $\bigtriangledown$} represent the mass distribution for $t$
$=$ $112$ $s$, $355$ $s$, $1122$ $s$ and $3550$ $s$, respectively.
$z$ $=$ $0$ corresponds to the prism bottom and $z$ $=$ $1$ to the
top. The dashed lines indicates the boundaries of the three
defined regions.}

\end{figure}

Figure \ref{fig4} shows the time evolution of the cluster size
distribution and weight average cluster size for the bottom region
of the prism. Since this subsystem gains mass from the upper
regions, its particle concentration continuously increases and
hence, some remarkable differences appear. The most important is
that the average $n_w$ becomes a monotonously increasing function
of time. Moreover, $n_w$ seems to follow an asymptotical power
behavior, $n_w$ $\sim$ $t^k$, where $k$ $\simeq$ $4$. Due to the
finite extension of the studied system, the asymptotical behavior
cannot be prolonged in time and hence, $n_w$ diminishes its rate
of change until the remaining clusters settle and become part of
the sediment. Another difference is that larger clusters are
formed at the bottom faster than in the bulk. Furthermore, they
also react faster to form the sediment and so, their time
evolution shows a double peak. The former peak (the smallest)
corresponds to those clusters aggregating at the bottom to form
the sediment and the second to the bulk clusters. Monomers also
behave differently. During the final stages their concentration
abruptly change from diminishing to increasing. This is a
consequence of the monomer flow coming from the middle region that
prevails over the small remaining aggregation rate. For even
longer times, monomers eventually collide with the sediment and
hence, they finally disappear.

\begin{figure}
\resizebox{0.465\textwidth}{!}{\includegraphics{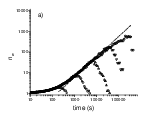}}
\resizebox{0.465\textwidth}{!}{\includegraphics{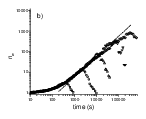}}
\resizebox{0.465\textwidth}{!}{\includegraphics{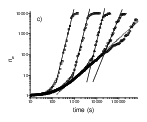}}
\caption{\label{fig6} Time evolution of the weight average cluster
size, $n_w$, for different Peclet numbers, $Pe$, and for the upper
a), middle b), and bottom c), portions of the system. The symbols
{\tiny $\Box$}, $\circ$, {\tiny $\bigtriangleup$}, {\tiny
$\bigtriangledown$} and $\diamond$ correspond to the evolutions
obtained under $Pe$ $=$ $1$, $0.1$, $0.01$, $0.001$ and $0.0001$,
respectively. The dashed lines represent the asymptotical DLCA
behavior. In figure c), the continuous lines are drawn as a guide
to the eye.}
\end{figure}

In order to study the mass distribution profile along the
sedimentation direction, $z$, the system was subdivided in 20
slices. For each one and for a given time, its total mass was
calculated by $m(z)$ $=$ $ \left. \sum_iin\right|_z$ and
represented as a point in figure \ref{fig5}. This figure was
constructed for $Pe$ $=$ $0.1$, i.~e.~corresponds to the data
shown in figures \ref{fig2}, \ref{fig3} and \ref{fig4}. It is
observed for time $112$ $s$ that the mass distribution is
practically uniform. Only for the upper and bottom slices, slight
deviations from the mean value are observed. While the system
evolves in time, those deviations becomes larger due to the
settling process. At $355$ $s$, the mass concentration of the
upper slice is about one half of the initial value and the mass
concentration of the bottom slice is approx.~the double. The other
slices, however, keep their initial mass concentration. After a
sufficiently long time (see $1122$ $s$), no remaining particles
are observed in the upper slice and the mass concentration
diminishes strongly in the two contiguous slices. Nevertheless,
the other slices do not change their mass concentration except for
the bottom one, which contains the sediment. This situation
changes for very long times, where the initial mass concentration
is not maintained any more in any slice of the system. This can be
seen for time $3550$ $s$. Here, the sediment contains more than
half of the total mass. It should be noted that the mass
concentration is not larger than the initial concentration except
for the bottom slice. This means that the assumption made in
section \ref{simul} of neglecting the backflow effect and the
hydrodynamic forces due to the low concentration is valid for
almost the whole system.

The time evolution of the weight average cluster size, $n_w$, for
the Peclet numbers $Pe$ $=$ $1$, $0.1$, $0.01$, $0.001$ and
$0.0001$ is shown in figure \ref{fig6}. Again, three different
plots are employed for representing the evolution of each region.
In addition, all figures include a unity sloped straight line,
which represents the asymptotic evolution of $n_w$ for DLCA. As
expected, the sedimentation effects appears at shorter times as
the Peclet number is increased. This is clearly seen for the three
prism regions. On the other hand, the beginning of the coupled
aggregation and sedimentation process evolves similarly to one
following a pure DLCA regime for all the studied Peclet numbers.
This indicates that a very large $Pe$ is needed for appreciating
the sedimentation effects at the beginning of the processes.
Figures \ref{fig6} a) and b) show that the DLCA asymptotic
evolution is, generally, not surpassed by the curves that consider
sedimentation. In fact, when settling effects appear the weight
average cluster size starts diminishing. Furthermore, it is
observed for all Peclet numbers that the $n_w$ curves peak earlier
and reach lower values for the upper region. On the contrary, in
figure \ref{fig6} c) it is observed that the curves behaves
completely different. Here, the sedimentation effects also change
the $n_w$ evolution by strongly increasing its slope. We found for
$Pe$ $=$ $1$, $0.1$ and $0.01$ a practically constant slope of
approximately 4, which is, by far, larger than the DLCA asymptotic
slope. The continuous lines shown in figure \ref{fig6} c) for this
Peclet numbers were drawn by considering this slope. For $Pe$ $=$
$0.001$ a slope of approx. 2.5 was obtained and for $Pe$ $=$
$0.0001$ no significant deviation from the pure DLCA slope was
observed. It is likely that, at least for our simulation
conditions, a limiting value of 4 is achieved for the $n_w$
asymptotic slope.

\begin{figure}
\resizebox{0.47\textwidth}{!}{\includegraphics{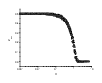}}
\caption{\label{fig7} a) $z_{cm}$ as a function of $\theta$ for
$Pe$ $=$ $1$ ({\tiny $\Box$}), $0.1$ ($\circ$), $0.01$ ({\tiny
$\bigtriangleup$}), $0.001$ ({\tiny $\bigtriangledown$}) and
$0.0001$ ($\diamond$), represented in a semilog plot.}
\end{figure}

The vertical position of the system center of mass, $z_{cm}$, is
plotted against the dimensionless time $\theta$ $=$ $t
(Pe)^{\alpha}/t_{ag}$ in figure \ref{fig7}, for the set of Peclet
numbers. Here, $t_{ag}$ $=$ $2/C_0K_{11}^{Smol}$ $=$ $352 s$ is
the characteristic aggregation time, $C_0$ $=$ $5.12$ $\times$
$10^{14}$ $m^{-3}$ is the initial particle concentration,
$K_{11}^{Smol}$ $=$ $8k_BT/$ $(3\eta)$ is the dimer formation
Smoluchowski's rate constant, which takes the value $11.1$
$\times$ $10^{-18} m^3 s^{-1}$ for the assumed solvent conditions.
In addition, the parameter $\alpha$ was introduced in order to
attempt to define a single master curve for the complete set of
curves. As figure \ref{fig7} clearly shows, the master curve is
yielded for $\alpha$ $=$ $0.74$. This curve takes a value closed
to $0.5$ at the beginning of the processes since the mass is
randomly scattered at time zero for each system, $z_{cm}$. At
first stages and for all $Pe$, $z_{cm}$ does practically not
evolve in time. Once the aggregation process leads to large enough
clusters, the settling process starts taking place. This starting
point strongly depends on the Peclet number. The larger the Peclet
number is, the sooner the settling process appears since a smaller
cluster size is needed for obtaining an appreciably sedimentation
velocity. Once the settling effects appear, the system center of
mass moves quickly towards the prism base until the mass transfer
from dispersion to sediment is finished.

\begin{table*}
\caption{\label{table1} Cluster fractal dimensions, $d_f|_{h}$,
and average ratios $\langle r_{gxy}\rangle / \langle r_{gz}\rangle
|_h$ as a function of the Peclet number. $h$ refers to the $top$,
the $middle$ and the $bottom$ regions. The last column indicates
whether the system percolates at the prism base.}
\begin{tabular}{c|ccccccc}
$Pe$ & $d_f|_{top}$ & $d_f|_{middle}$ & $d_f|_{bottom}$ & $\langle
r_{gxy}\rangle / \langle r_{gz}\rangle |_{top}$ & $\langle
r_{gxy}\rangle / \langle r_{gz}\rangle |_{middle}$ & $\langle
r_{gxy}\rangle /
\langle r_{gz}\rangle |_{bottom}$ & percolation  \\
\hline $1$ & $--$ & $--$ & $1.67\pm0.05$& $1.01\pm0.01$ & $0.99\pm0.03$ & $0.90\pm0.02$ & Yes \\
$0.1$ & $1.88\pm0.07$ & $1.89\pm0.06$ & $1.74\pm0.05$ & $1.00\pm0.05$ & $1.01\pm0.02$ & $0.95\pm0.02$ & Yes\\
$0.01$ & $1.80\pm0.06$ & $1.83\pm0.06$ & $1.82\pm0.05$ & $0.99\pm0.05$ & $1.00\pm0.02$ & $0.99\pm0.02$ & Yes \\
$0.001$ & $1.82\pm0.06$ & $1.84\pm0.06$ & $1.81\pm0.05$ & $1.02\pm0.05$ & $1.00\pm0.02$ & $1.01\pm0.02$ & No \\
$0.0001$ & $1.83\pm0.06$ & $1.80\pm0.06$ & $1.82\pm0.05$ & $1.00\pm0.05$ & $0.99\pm0.02$ & $1.00\pm0.02$ & No\\
$0$ & $1.78\pm0.06$ & $1.74\pm0.06$ & $1.76\pm0.05$ & $0.99\pm0.05$ & $1.00\pm0.02$ & $1.00\pm0.02$ & No \\
\end{tabular}
\end{table*}

As explained in section \ref{simul}, the cluster fractal dimension
was assessed by means of the radii of gyration method and
calculated for each defined region. For $Pe$ $=$ $1$ and for the
upper regions, it was not possible to obtain a reliable fractal
dimension value since not enough large clusters were formed in
these regions. For obtaining a reliable statistics, three
simulation runs were performed for each Peclet number. Pure DLCA
simulations were also carried out as a reference. The results are
shown in table \ref{table1}. For $Pe$ $=$ $0$, i.~e.~pure DLCA,
the values are consistent with the ones reported in the literature
\cite{broide90,jcolloid01}. Furthermore, no significant difference
appears among the regions, as expected for a regime where no
sedimentation velocity was imposed. For larger values of the
Peclet number it is observed that the cluster fractal dimension
increase for the top and middle regions. This is in agreement with
the simulations results obtained by Gonzalez, in which no extra
effects are needed, such us cluster restructuring, for explaining
the increase of the fractal dimension \cite{gonzalez01}.
Nevertheless, Gonzalez obtained $d_f$ $=$ $2.27$ for the larger
structures and for $Pe$ $=$ $0.01$, which is larger than our
results. This may be a consequence of the higher concentrations
employed in his simulations as discussed further in the text. In
addition, the $d_f|_{middle}$ values seems to be slightly larger
than the $d_f|_{top}$ values. This may be due to the average
downward distance that the corresponding clusters moved. Since
this distance is longer for the clusters inside the middle region,
the effect of increasing the fractal dimension due to the Stokes
velocity becomes more evident.

\begin{figure}
\resizebox{0.47\textwidth}{!}{\includegraphics{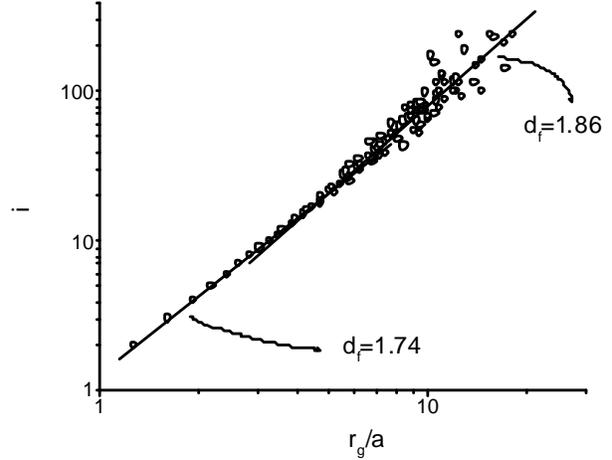}}
\caption{\label{fig8} The cluster size, $i$, as a function of the
normalized radius of gyration, $r_g/a$, for $Pe$ $=$ $0.01$ and
for the middle region. The solid lines represents the linear fits
obtained for the ranges $r_g/a$ $\epsilon$ $[2,6]$ and $r_g/a$
$\epsilon$ $[6,18]$, respectively. }
\end{figure}

For $Pe$ $=$ $0.01$ and for the middle region, figure \ref{fig8}
shows the cluster size, $i$, as a function of the normalized
radius of gyration, $r_g/a$. The data were fitted over two ranges
and different slopes were obtained. This means that a crossover
between the two different power lows may, at least, not be
discarded. This behavior was already reported in ref.
\cite{gonzalez01} where it is explained that smaller clusters show
a roughly DLCA behavior, whereas larger ones behave differently
since they are more influenced by the external field. Furthermore,
it is expected that the crossover cluster size decreases as $Pe$
increases as reported in ref. \cite{gonzalez01}. This is not easy
to see in our simulations since the clusters disappear from the
middle region before growing enough, and this leads to poorer
statistic results. It should be pointed out that table
\ref{table1} only reports average values (considering only those
clusters containing more than 15 particles), which were employed
to dictate the cluster motion, i.~e. no distinction was made
between the fractal dimensions of small and large clusters.

The fractal dimensions obtained for the bottom region varies in a
quite different way with the Peclet number than for the upper
regions. Here, it is observed that $d_f$ strongly decreases as the
Peclet number increases. It was expected, however, to obtain even
larger fractal dimensions than the ones obtained in the upper
regions. The fact of obtaining a fractal dimension as low as $d_f$
$=$ $1.68\pm0.05$ for $Pe$ $=$ $1$ clearly indicates that the
sediment formation is changing in some way the growth process.
Since the cluster movement is restricted almost to two dimensions
at the prism base and since the fractal dimension yield for DLCA
processes at two dimensions is $d_f$ $=$ $1.45\pm0.05$, it is not
very surprising to find cluster structures characterized by
fractal dimensions ranging between $1.75$ and $1.45$
\cite{Marmur79,europhysicsJ01}. This indicates that for increasing
values of the Peclet number, the average cluster size of the
settling aggregates that arrive to the prism bottom decreases and
hence, most clustering reactions take place in the bottom region.

The average ratio $\langle r_{gxy}\rangle$/$\langle r_{gz}\rangle
|_h$ $=$ $(\langle r_{gx}\rangle$/$\langle r_{gz}\rangle|_h$ $+$
$\langle r_{gy}\rangle$ /$ \langle r_{gz}\rangle|_h)/2$ obtained
for different Peclet numbers is shown in table \ref{table1}. As
explained in section \ref{simul}, this quantity accounts for the
average shape of the clusters. In case that $\langle
r_{gxy}\rangle / \langle r_{gz} \rangle$ $>$ $1$, the clusters
tend to be elongated in the $z$ direction and when $\langle
r_{gxy}\rangle / \langle r_{gz} \rangle$ $<$ $1$ the clusters are
shorter. For the upper regions, no significative deviation from
unity is observed not even for the higher Peclet numbers. This
indicates that the clusters do not show a preferential growth
direction. This is in good agreement with Allain
et.~al.~experimental observations \cite{Allain95,Allain96}.
Nevertheless, our results do not agree with Gonzalez findings
\cite{gonzalez01}. For Peclet numbers as high as 0.1 and for much
more concentrated systems ($\phi$ $=$ $0.01$), Gonzalez found that
clusters grow faster in the sedimentation direction. This is very
likely to occur since clusters move mostly downwards. For diluted
systems, on the other hand, when the Peclet number is as high as
$0.1$, the clusters settle so fast that they have not the
opportunity to grow before arriving to the prism base. This fact
explains why we do not obtain $\langle r_{gxy}\rangle / \langle
r_{gz} \rangle$ significatively larger than unity for the upper
regions. The same argument is also valid to explain the
differences between the fractal dimensions obtained by Gonzalez
and by us. For the lower region and where the sediment grows, for
the highest Peclet numbers we obtain $\langle r_{gxy}\rangle /
\langle r_{gz} \rangle$ lower than unity. This is a consequence of
the fact that aggregation is taking place mostly at the prism
bottom, at which the $z$ direction motion is restricted. Hence,
the sediment grows covering the base of the prism producing local
two dimensional percolation. This kind of percolation, as shown in
table \ref{table1}, occurs only for the largest values of the
Peclet number.

\begin{figure}
\resizebox{0.47\textwidth}{!}{\includegraphics{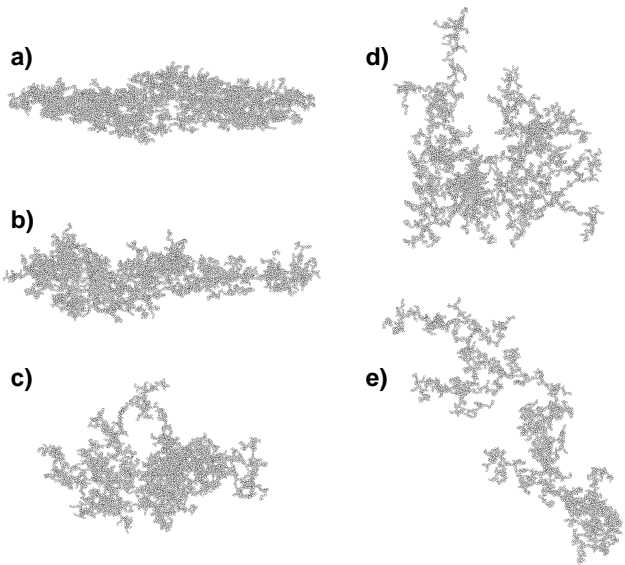}}
\caption{\label{fig9} Three-dimensional perspective plot of the
sediments obtained for $Pe$ $=$ 1 a), 0.1 b), 0.01 c), 0.001 d)
and 0.0001 e). No sediment was obtained for $Pe$ $=$ $0$.}
\end{figure}

\begin{figure}
\resizebox{0.47\textwidth}{!}{\includegraphics{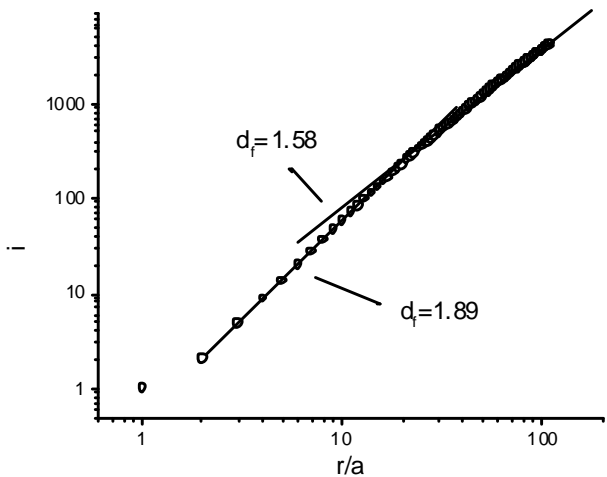}}
\caption{\label{fig10} The $i(r/a)$ function averaged for the
three sediments obtained under $Pe$ $=$ $1$. }
\end{figure}

Finally, figure \ref{fig9} shows the sediments obtained for
different Peclet numbers. It can clearly be seen that for
increasing $Pe$ the sediments become more extended in the base
plane of the prism and show a higher degree of compactness. For
instance, sediments a) and b) almost cover the prism bottom
completely. In order to study the structure of the sediment, a
fractal analysis was realized for each sediment cluster. For this
purpose, the distances between the centers of each particle and
the center of a given reference particle are calculated. This
allows to build a function $i(r)$, where $i$ denotes the number of
particles inside a sphere defined by the reference particle center
and the radius $r$. The procedure is repeated by changing the
reference particle until all particles had been considered as
reference, while averaging all obtained $i(r)$ functions. Then,
$d_f$ may by obtained from the relationship $r$ $\sim$
$i^{1/d_f}$. For improving the statistics, three sediments were
analyzed for each $Pe$ and the corresponding $d_f$ values were
averaged. This procedure yields well defined $d_f$ for the smaller
Peclet numbers, having $d_f$ $=$ $1.82\pm0.05$, $1.91\pm0.06$ and
$1.90\pm0.06$ for $Pe$ $=$ $0.0001$, $0.001$ and $0.01$,
respectively. This values are all slightly higher than the
corresponding values shown in table \ref{table1}, which may be due
to differences on the applied methods. Anyway, the same tendency
for increasing $Pe$ is verified. For $Pe$ $=$ $0.1$ and $1$, two
different power low behaviors were found for the $i(r)$ functions.
A short range $d_f$ $=$ $1.99\pm0.07$ and $1.89\pm0.07$ and a long
range $d_f$ $=$ $1.58\pm0.07$ and $1.67\pm0.07$ were obtained for
$Pe$ $=$ $0.1$ and $1$, respectively. This points towards a
crossover from an aggregation mechanism which takes place mostly
in the bulk to another aggregation mechanism restricted almost to
two dimensions, i.~e. processes occurring at the prism bottom.
Figure \ref{fig10} shows the averaged $i(r/a)$ function for $Pe$
$=$ $1$.

\section{\label{conclusions}Conclusions}

Coupled aggregation and sedimentation process were simulated by
considering a large prism with no periodical boundary conditions
for the sedimentation direction. Three equally sized and mutually
excluded regions were defined for studying their time evolution of
the cluster size distribution. We found that for the upper regions
the shape of the time evolution of oligomers narrows due to the
effect of the Stokes velocity. On the contrary, the population
dependence on time shows a double peak for the bottom region as a
consequence of the settling clusters that arrive at the prism
base. The time evolution of the weight average cluster size also
changes its behavior depending on the region. While it shows a
peak for the upper regions, it becomes a monotonously increasing
function for the bottom region. Furthermore, the limiting value of
$4$ was obtained for its slope for increasing values of the Peclet
number.

The cluster structure was also studied by means of measuring the
cluster radii of gyration. In agrement with the experiments
performed by Allain et.~al., we obtained that there is not a
preferential growth direction. We also found an increase in the
fractal dimension for increasing Peclet number in the upper
regions. For the bottom region, however, a decrease in the fractal
dimension and the preferential growth directions parallel to the
prism base were found for increasing Peclet number. In addition,
for the highest $Pe$, the sediment fractal dimension crosses over
from high values, corresponding to the smaller radii, to values
close to $1.45\pm0.05$, for larger radii, which is the accepted
value for two dimensional DLCA. These facts are explained as a
consequence of the restricted motion of the settling clusters that
have already arrived at the prism base. Finally, one may conclude
that increasing the Peclet number causes the system to increase
its tendency to produce a two dimensional percolation at the prism
base.

\end{document}